\documentstyle[psfig]{mn}

\newif\ifAMStwofonts

\def\lsimeq{{_<\atop^{\sim}}}
\def\gsimeq{{_>\atop^{\sim}}}
\def\oii{[O{\tt II}]}
\def\oiii{[O{\tt III}]}
\def\mgii{Mg{\tt II}}
\def\ciii{C{\tt III}]}
\def\cii{C{\tt II}]}
\def\civ{C{\tt IV}}
\def\lya{Ly$\alpha$}
\def\hal{H$\alpha$}
\def\hbe{H$\beta$}
\def\eso36{\hbox{ESO~3.6--m}}
\title[Optical identifications of faint radio sources]
{Optical identifications and spectroscopy of a faint radio source sample:
the nature of the sub--mJy population\thanks{Based on observations collected 
at the European Southern Observatory, LaSilla, Chile}}
\author[C. Gruppioni, M. Mignoli \& G. Zamorani]
{C. Gruppioni$^{1}$
, M. Mignoli$^{2}$ \& 
G. Zamorani$^{2,3}$ \\
$^1$ Imperial College of Science, Technology and Medicine, Prince
    Consort Road, London SW7 2BZ, U.K.\\
$^2$ Osservatorio Astronomico di Bologna, via Zamboni 33, I--40126
    Bologna, Italy \\
$^3$ Istituto di Radioastronomia del CNR, via Gobetti 101, I--40129
    Bologna, Italy \\
e-mail: c.gruppioni@ic.ac.uk
}
\date{Accepted 1998 November 18. Received 1998 September 24}
\pagerange{\pageref{firstpage}--\pageref{lastpage}}
\pubyear{1997}

\def\LaTeX{L\kern-.36em\raise.3ex\hbox{a}\kern-.15em
    T\kern-.1667em\lower.7ex\hbox{E}\kern-.125emX}

\begin{document}
 
\label{firstpage}

\maketitle
 
\begin{abstract}
Deep imaging and spectroscopy have been carried out for optical counterparts
of a sample of 68 faint radio sources ($S > 0.2$ mJy) in the ``Marano Field''.
About 60\% of the sources have been optically identified on deep CCD 
exposures (limit $R \simeq 24.0$) or \eso36 \ plates (limit $b_J \simeq 22.5$).
Thirty--four spectra (50\% of the total radio sample) were obtained with
the \eso36 telescope and 30 redshifts were determined. In addition to a few broad 
line active galactic nuclei,
three main spectroscopic classes have been found to dominate the faint radio
galaxy population: (1) Early--type galaxies
(without emission--lines in their spectra) having $0.1 < z < 0.8$ and covering 
the range of radio fluxes 0.2 -- 30 mJy. (2) Late--type galaxies (with narrow 
emission--lines in their spectra) at moderate redshift 
($z < 0.4$), with relatively bright magnitudes ($B < 22.5$) and sub-milliJanski radio fluxes. 
When applicable, the diagnostic diagrams for these sources are consistent with the lines 
being produced by star--formation activity. (3) A group of bright high--redshift ($z > 0.8$) 
radio galaxies with  moderate--to--strong \oii \ emission. All of them have $B > 22.5$
and most of them have $S_{1.4~GHz} > 1$ mJy. They have spectra, colours and absolute
magnitudes similar to those of the classical bright elliptical radio galaxies found in
surveys carried out at higher radio fluxes. Star--forming galaxies do not constitute
the main population of our radio sources identified with galaxies. In fact, even at sub--mJy
level the majority of our radio sources are identified with early--type galaxies. This 
apparent discrepancy with previous results is due to the fainter magnitude limit reached
in our spectroscopic identifications. Moreover, using mainly the large radio--to--optical 
ratio and the information from the available limits on the optical magnitudes of the
unidentified radio sources, we conclude that the great majority of them are likely
to be early--type galaxies, at $z > 1$. If correct, it would suggest that the evolution
of the radio luminosity function of spiral galaxies, including starbursts, might not
be as strong as suggested in previous evolutionary models.
\end{abstract}

\begin{keywords}
galaxies: active -- galaxies: starburst -- cosmology: observations -- 
radio continuum: galaxies.
\end{keywords}

\section{Introduction}
In the last decade, deep radio surveys (Condon \& Mitchell 1984; Windhorst
1984; Windhorst et~al. 1985) have pointed out the presence of
a new population of radio sources appearing below a few mJy and
responsible for the observed flattening in the differential
source counts (normalized to Euclidean ones).
Several scenarios have been developed to 
interpret this ``excess'' in the number of faint radio sources,
invoking a non--evolving population of local ($z < 0.1$) low--luminosity
galaxies (Wall et~al. 1986), strongly--evolving normal spirals (Condon 1984,
1989) and actively star--forming galaxies (Windhorst et~al. 1985, 1987;
Danese et~al. 1987; Rowan--Robinson et~al. 1993). The latter scenario is
supported by the existing optical identification works performed for the 
sub--mJy population. These works have, in fact,
shown that the sub--mJy sources are mainly identified with faint blue 
galaxies (Kron, Koo \& Windhorst 1985; Thuan \& Condon 1987), often showing
peculiar optical morphologies indicative of interaction and merging phenomena
and spectra similar to those of the star--forming galaxies detected by IRAS
(Franceschini et~al. 1988; Benn et~al. 1993). However, since the majority of
these objects have faint optical counterparts, visible only in deep CCD
exposures (down to $B \sim$ 24--25), all these works are based on small
percentages of identification. For example, the Benn et~al. spectroscopic
sample, despite the fact that it is the largest sample so far available in
literature, corresponds to slightly more than 10 per cent of the total radio
sample. 

In order to better understand the nature of the sub--mJy radio galaxy 
population on the basis of a larger identification fraction than the ones 
obtained so far, we performed deep photometric and spectroscopic
identifications for a faint radio source sample in the ``Marano Field''.
Here we present the results of the identification of 68 objects, which
represent the total radio sample obtained by joining together the two $S > 0.2$
mJy complete samples at 1.4 and 2.4 GHz in the ``Marano Field''
(see Gruppioni et~al. 1997). We were able to reach a relatively high 
identification fraction with respect to previous works, since we optically
identified 63\% of the 68 radio sources and we obtained spectra for 34 of them
($\sim$50\% of the total sample). These constitute the highest identification
fractions so far available in literature for sub--mJy radio samples.

The paper is structured as follows:
in section 2 we describe the radio sample; in section 3 we present the 
photometric data and optical identifications; in section 4 we present the
spectroscopic results, including spectral classification for the optical
counterparts and notes on individual objects;
in section 5 we discuss the radio and optical properties of the faint radio
source population; in the last two sections we discuss our results and
present our conclusions.    
  
\section{The Radio Sample}
Deep radio surveys with the Australia Telescope Compact Array (ATCA) have been
carried out at 1.4 and 2.4 GHz, with a limiting flux of $\sim$0.2 mJy at each
frequency, in the ``Marano Field'' (centered at $\alpha(2000) = 03^h 15^m 09^s,
\delta(2000) = -55^{\circ} 13^{\prime} 57^{\prime \prime}$), for which deep
optical and X--ray data are also available.\\
The two radio samples, complete at the 5$\sigma_{local}$ level, consist of 63
and 48 sources respectively at 1.4 and 2.4 GHz. The main results of the
radio data analysis are extensively described by Gruppioni et~al. (1997).
The 1.4 GHz differential source counts show the flattening below about 1 mJy
found by previous authors (Condon \& Mitchell 1984; Windhorst et~al. 1985) and
considered as the typical feature of the sub--mJy population. 
From the study of the spectral index distribution as a function of flux, a 
significant flattening of the spectral index toward fainter fluxes has been
found for the higher frequency selected sample (2.4 GHz), while the median
spectral index ($\alpha_{med}$) is consistent with remaining constant at $\sim
$0.8 ($f_{\nu} \propto \nu^{-\alpha}$) for the sample selected at 1.4 GHz. 
However, at both frequencies a significant number of sources with inverted
spectrum do appear at flux densities $\lsimeq 2$ mJy. In particular, objects 
with inverted spectra constitute $\sim$13\% of the total 1.4 GHz sample and
$\sim$25\% of the total 2.4 GHz one. For the latter sample this percentage 
increases to $\sim$40\% for $S < 0.6$ mJy. 

The total radio source sample considered in this paper for the optical 
identification work consists of the 1.4 GHz complete sample joined together
with the 5 sources detected only at 2.4 GHz above the 5$\sigma_{local}$ level.

\section{Photometry and Optical Identification}
For the identification of our radio sources we used the photometric
data already available for the ``Marano Field''. These data consist 
of \eso36 \ plates in the bands $U_K$, $J_K$ and $F_K$ (Kron 1980)
covering an area of $\sim$0.69~sq.deg. (the plates include the entire radio
field, that is $\sim$0.36 sq. deg.) and reaching limit in magnitudes of 
\hbox{$J_K \sim$ 22.5} and \hbox{$U_K,F_K \sim$ 22.0}. The data from the plates
have been utilized for the selection and definition of a complete sample of
faint quasars with $J \leq 22.0$, the MZZ sample (Marano, Zamorani 
\& Zitelli 1988; Zitelli et~al. 1992). Moreover, the inner part of the field
($\sim$15$^{\prime}$ radius) has been observed with the ESO NTT telescope,
with deep CCD exposures in the $U$, $B$, $V$ and $R$ bands, down to 
limit in magnitudes $U$$\sim$23.5, $B$$\sim$25.0, $V$$\sim$24.5 and $R$$\sim$24.0.
With successive observing runs a rather complete coverage of the central
part of the field (corresponding to $\sim$60\% of the area covered by the radio
observations) has been obtained in the $V$ and $R$ bands (see Figure~1). 

\begin{figure}
\centerline{
  \psfig{figure=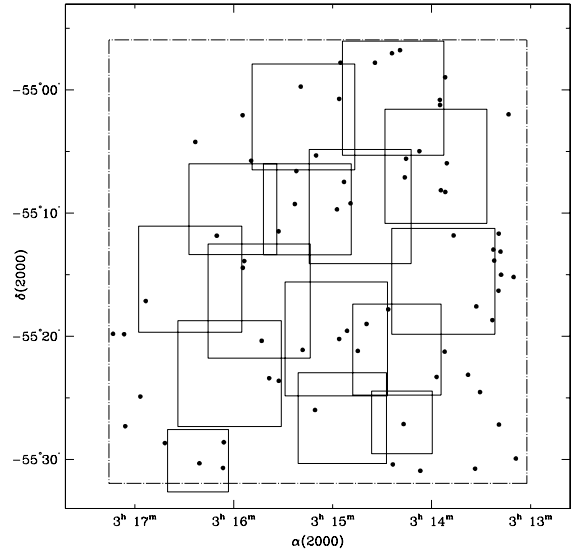,width=8cm}
}
 \caption{\label{fig1} CCD coverage of the inner part of the ``Marano Field''
in the $V$ and $R$ bands. The area covered by 
the radio observations is represented by the dashed box. The filled dots are the 68 
radio sources.} 
\end{figure}

A multi--colour catalog based on the deep CCD observations has been created.
The images were reduced in a standard way using {\tt IRAF}\footnote{{\tt IRAF}
is distributed by the National Optical Astronomy Obser\-vatories, which are
operated by AURA Inc. for the NSF.} and for the data analysis
(detection, photometry and star/galaxy classification)
we used SExtractor (Bertin and Arnouts 1996). More details on the data
reduction and analysis can be found elsewhere (Mignoli, 1997) and a full
description of the catalog is in preparation. 
We used the CCD overlapping regions to check the photometric errors and
homogeneity in the detection and morphological classification.
In the B band the typical errors are $\sim$0.05 mag up to
$B = 23$, growing to 0.15$\div$0.20 at the limiting magnitude of B=25.
In the $R$ band the error ranges from $\sim$$0.07$ at $R = 22$ to $\sim$$0.20$ at 
$R = 24$. 

For the identification of the 50 radio sources which
fall in the $R$ band CCD frames, we used the magnitudes and
positions of the optical sources given in the catalog.
For the 18 radio sources outside the area covered
by CCD frames, we used the magnitudes and positions of the optical
sources taken from the plates data. In this case the $U_K$, $J_K$ and $F_K$
magnitudes of the photographic system have been transformed to $U$, $B$, $V$ 
and $R$ magnitudes of the standard Johnson/Cousin system of the CCD data,
according to the transformation formulae given by Kron (1980) and Koo \& Kron
(1982): 
$$ U \approx U_K \qquad {V+R\over2} \approx F_K $$
$$ B = J_K + 0.20(J_K - F_K) $$ 
$$ V = F_K + 0.34(J_K - F_K) $$
The correctness of this transformation between the two photometric systems 
has been checked using the
objects detected both in the plates and in the CCD frames: in all the four
bands the average of the magnitude differences is less than 0.05 mag,
implying a good consistency between the CCD and the plates photometry,
whereas the scatter of the points gives an estimate of the random errors of 
$\sim$0.10~mag for the $B,V$ bands and of $\sim$0.15~mag for the $U,R$ bands. 

\begin{table*}
 \centering
 \begin{minipage}{150mm}
  \caption{Optical Identifications}
\begin{tabular}{ccrrrrrrrrrc}
 & & & & \\ \hline \hline
 & & & & & \\
\multicolumn{1}{c}{N$_{1.4}$} & \multicolumn{1}{c}{N$_{2.4}$} & 
\multicolumn{1}{c}{$S_{1.4}$} & \multicolumn{1}{c}{$S_{2.4}$} & 
\multicolumn{1}{c}{$\alpha_{r}$} &
\multicolumn{1}{c}{$U$} & \multicolumn{1}{c}{$B$} & \multicolumn{1}{c}{$V$} &
\multicolumn{1}{c}{$R$} & \multicolumn{1}{c}{$\Delta$} &
\multicolumn{1}{c}{$L$} & \multicolumn{1}{c}{Envir} \\
 & & & & \\
 & & (mJy) & (mJy) & & & & & & ($^{\prime \prime}$) & & \\
 & & & & \\ \hline
 & & & & \\
   01 & 01 &    8.28  &    4.86  &     0.97  &         &$>$~22.5*&         &$>$~21.8*&     &      &     \\ 
   02 &$--$&    0.62  & $<~$0.53 & $>$~0.28  &         &$>$~22.5*&         &$>$~21.8*&     &      &     \\
   03 & 02 &   26.40  &   26.14  &     0.02  &  19.5*  &  19.3*  &  18.0*  &  17.1*  & 1.5 & 58.9 & GR  \\ 
   04 &$--$&    0.67  & $<~$0.48 & $>$~0.62  &         &         &$>$~24.50&$>$~24.00&     &      &     \\ 
   05 & 03 &    0.54  &    0.79  &    -0.68  &         &         &$>$~24.50&  23.21  & 1.0 &  5.7 &~GR? \\
      &    &          &          &           &         &$>$~22.5*&  23.34  &  22.03  & 2.1 &  4.5 &     \\ 
   06 &$--$&    0.78  &    0.40  &     1.21  &  17.6*  &  17.9*  &  17.0*  &  16.9*  & 2.4 & 28.8 &     \\ 
   07 & 04 &    0.80  &    1.01  &    -0.42  &  21.4*  &  21.6*  &  21.04  &  20.60  & 3.2 &  1.8 &     \\ 
   08 &$--$&    0.39  & $<~$0.28 & $>$~0.60  &  19.2*  &  19.9*  &  19.66  &  19.43  & 0.3 & 52.4 &     \\ 
   09 & 05 &    1.52  &    2.21  &  $-$0.68  &  20.4*  &  19.8*  &  18.34  &  17.65  & 1.1 & 57.2 &     \\ 
  $--$& 06 & $<~$0.32 &    0.46  & $<-$0.66  &         &         &$>$~24.50&  24.60  & 1.4 & $--$ &     \\ 
   10 & 07 &   20.02  &   15.90  &     0.42  &         &         &$>$~24.50&$>$~24.00&     &      &     \\ 
   11 &$--$&    0.49  & $<~$0.50 & $>-$0.03  &  21.6*  &  22.2*  &  21.4*  &  20.9*  & 2.8 &  2.2 &     \\ 
   12 &$--$&    0.50  &    0.32  &     0.83  &$>$~23.50&$>$~24.50&$>$~24.50&$>$~24.00&     &      &     \\ 
   13 & 08 &    0.60  &    0.51  &     0.28  & 19.8*   &  20.0*  &  19.1*  &  18.8*  & 0.6 & 71.0 &     \\ 
   14 & 09 &    2.91  &    1.60  &     1.09  &         &$>$~22.5*&         &$>$~21.8*&     &      &     \\ 
   15 & 10 &  158.00  &   95.94  &     0.90  & 19.39   &  19.91  &  19.77  &  19.41  & 1.1 & 40.3 &     \\ 
  $--$& 11 &    0.20  &    0.30  &  $-$0.72  &         &         &$>$~24.50&$>$~24.00&     &      &     \\ 
   16 & 13 &    1.33  &    0.83  &     0.85  &         &         &$>$~24.50&$>$~24.00&     &      &     \\ 
   17 & 12 &    3.08  &    1.76  &     1.02  &         &  23.1*  &  22.3*  &  21.7*  & 1.1 &  4.9 &     \\ 
   18 &$--$&    0.28  & $<~$0.20 & $>$~0.64  & 21.0*   &  20.9*  &  19.7*  &  19.2*  & 1.6 & 44.1 & ~D~ \\
   19 & 14 &    6.27  &    5.31  &     0.30  &$>$~23.50&$>$~24.50&$>$~24.50&$>$~24.00&     &      &     \\
   20 & 15 &    0.38  &    0.43  &  $-$0.24  &         &$>$~22.5*&  23.50  &  22.12  & 0.9 &  9.9 & CL  \\
   21 & 16 &    1.02  &    1.00  &     0.04  &         &        &$\sim$25.0&$\sim$24.0&1.0 & $--$ &     \\
   22 &$--$&    0.47  &    0.28  &     0.97  & 21.3*   &  22.2*  &  20.77  &  20.07  & 1.1 & 12.0 &     \\
   23 &$--$&    0.40  & $<~$0.24 & $>$~0.92  &         &$>$~22.5*&         &$>$~21.8*&     &      &     \\
   24 &$--$&    0.46  & $<~$0.36 & $>$~0.48  & 20.0*   &  20.4*  &  20.01  &  19.60  & 1.1 & 18.3 &     \\
   25 & 17 &    3.30  &    2.68  &     0.38  & 23.75   &  23.72  &  21.98  &  20.85  & 0.5 & 12.3 & ~D~ \\
      &    &          &          &           &$>$~23.50&  24.50  &  24.25  &  23.30  & 3.6 &  0.5 &     \\
      &    &          &          &           & 22.75   &  23.20  &  22.41  &  21.57  & 3.6 &  0.8 &     \\
  $--$& 18 &    0.20  &    0.33  &  $-$0.88  &         &$>$~22.5*&  23.85  &  23.72  & 4.8 &  0.0 &     \\
   26 & 19 &    0.45  &    0.48  &  $-$0.11  &         &  23.1*  &  21.6*  &  20.05  & 1.2 & 16.9 & ~D~ \\
   27 & 20 &    2.60  &    1.98  &     0.49  & 20.0*   &  19.6*  &  17.81  &  17.10  & 0.5 & 70.0 &     \\
   28 & 21 &    1.66  &    0.98  &     0.95  &         &$>$~22.5*&         &$>$~21.8*&     &      &     \\
   29 &$--$&    0.62  & $<~$0.44 & $>$~0.61  &         &         &$>$~24.50&$>$~24.00&     &      &     \\
   30 & 22 &    6.32  &    3.88  &     0.89  & 23.11   &  23.68  &  22.73  &  21.78  & 0.4 &  6.2 & GR  \\
      &    &          &          &           &$>$~23.50&  24.40  &  23.36  &  22.47  & 3.5 &  0.6 &     \\
   31 & 23 &    0.92  &    0.52  &     1.05  &         &$>$~22.5*&  22.78  &  21.58  & 2.4 &  2.7 & ~D~ \\
      &    &          &          &           &         &$>$~22.5*&  22.55  &  21.93  & 4.3 &  0.1 &     \\
   32 & 24 &    0.50  &    0.32  &     0.85  &$>$~23.50&  25.30  &  23.80  &  22.50  & 1.7 &  5.8 & GR  \\
      &    &          &          &           &$>$~23.50&$>$~24.50&  23.75  &  23.15  & 1.5 &  3.3 &     \\
   33 & 25 &    4.83  &    2.68  &     1.07  &$>$~23.50&$>$~24.50&$>$~24.50&$>$~24.00&     &      &     \\
   34 & 26 &    0.40  &    0.34  &     0.28  &$>$~23.50&$>$~24.50&$>$~24.50&  23.50  & 3.1 &  0.6 &     \\
   35 &$--$&    0.41  &    0.25  &     0.87  &$>$~23.50&  24.56  &  23.86  &  23.06  & 4.3 &  0.1 &     \\
   36 & 27 &    1.17  &    0.75  &     0.80  &         &        &$\sim$24.5&$\sim$24.3&1.4 &      &     \\
   37 & 28 &    1.01  &    0.62  &     0.87  &         &         &$>$~24.50&$>$~24.00&     & $--$ &     \\
   38 & 30 &    1.25  &    0.71  &     1.02  & 22.50   &  21.85  &  20.35  &  19.32  & 1.8 & 13.3 &     \\
   39 & 29 &   23.10  &   15.70  &     0.70  &         &         &$>$~24.50&$>$~24.00&     &      &     \\
   40 & 31 &    0.60  &    0.36  &     0.93  &$>$~23.50&$>$~24.50&$>$~24.50&$>$~24.00&     &      &     \\
   41 & 32 &    1.45  &    2.17  &  $-$0.73  &$>$~23.50&$>$~24.50&$>$~24.50&  23.92  & 2.0 &  3.1 &     \\
   42 & 33 &   30.12  &   19.74  &     0.77  &$>$~23.50&$>$~24.50&$>$~24.50&$>$~24.00&     &      &     \\
   43 & 34 &    0.42  &    0.32  &     0.47  & 20.73   &  20.58  &  19.30  &  18.60  & 1.4 & 47.1 &     \\
   44 & 35 &   15.08  &    9.59  &     0.82  &         &         &$>$~24.50&$>$~24.00&     &      &     \\
   45 &$--$&    0.20  &    0.21  &  $-$0.07  &$>$~23.50&  25.10  &$>$~24.50&$>$~24.00& 0.5 & $--$ &     \\
   46 &$--$&    0.50  & $<~$0.47 & $>$~0.11  &$>$~23.50&$>$~24.50&$>$~24.50&$>$~24.00&     &      &     \\
   47 &$--$&    0.47  &    0.27  &     0.98  &$>$~23.50&  22.86  &  21.39  &  20.05  & 1.1 & 16.2 & GR  \\
      &    &          &          &           & 21.50   &  21.86  &  21.38  &  20.75  & 4.8 &  0.1 &     \\
\hline 
\end{tabular}
\end{minipage}
\end{table*}

\begin{table*}
\centering
\begin{minipage}{150mm}
\contcaption{}
\begin{tabular}{ccrrrrrrrrrc}
 & & & & \\ \hline \hline
 & & & & & \\
\multicolumn{1}{c}{N$_{1.4}$} & \multicolumn{1}{c}{N$_{2.4}$} & 
\multicolumn{1}{c}{$S_{1.4}$} & \multicolumn{1}{c}{$S_{2.4}$} & 
\multicolumn{1}{c}{$\alpha_{r}$} &
\multicolumn{1}{c}{$U$} & \multicolumn{1}{c}{$B$} & \multicolumn{1}{c}{$V$} &
\multicolumn{1}{c}{$R$} & \multicolumn{1}{c}{$\Delta$} &
\multicolumn{1}{c}{$L$} & \multicolumn{1}{c}{Envir} \\
 & & & & \\
 & & (mJy) & (mJy) & & & & & & ($^{\prime \prime}$) & & \\
 & & & & \\ \hline
 & & & & \\
   48 & 36 &    0.54  &    0.34  &     0.82  & 19.01   &  18.98  &  17.96  &  17.40  & 0.6 & 92.6 &     \\
   49 & 37 &    9.04  &    6.66  &     0.55  &$>$~23.50&$>$~24.50&$>$~24.50&$>$~24.00&     &      &~CL? \\
   50 &$--$&    0.28  &    0.18  &     0.81  & 19.11   &  19.44  &  18.48  &  18.01  & 0.5 & 60.6 &     \\
   51 & 38 &    0.49  &    0.37  &     0.52  &         &$>$~22.5*&  23.10  &  21.90  & 0.5 &  9.5 & ~D~ \\
      &    &          &          &           &         &$>$~22.5*&  23.62  &  22.11  & 2.3 &  3.6 &     \\
   52 & 39 &    2.06  &    1.40  &     0.70  &         &$>$~22.5*&  23.20  &  22.70  & 1.5 &  5.4 &     \\
   53 & 40 &    1.54  &    1.79  &  $-$0.28  &$>$~23.50&$>$~24.50&  23.00  &  21.45  & 1.5 &  8.7 & CL  \\
   54 & 41 &    0.91  &    0.68  &     0.53  &         &$>$~22.5*&         &$>$~21.8*&     &      &     \\
   55 &$--$&    0.31  &    0.18  &     0.99  & 20.6*   &  20.4*  &  19.0*  &  18.22  & 1.2 & 38.3 & ~D~ \\
   56 & 42 &    2.38  &    1.87  &     0.44  &         &$>$~22.5*&         &  22.01  & 0.5 & 14.2 &     \\
   57 & 43 &    1.71  &    1.07  &     0.85  & 22.69   &  23.14  &  22.81  &  22.09  & 0.9 &  7.9 &     \\
   58 & 44 &    1.29  &    0.91  &     0.63  & 21.2*   &  21.3*  &  20.20  &  19.68  & 4.1 &  1.2 &     \\
      &    &          &          &           &$>$~23.50&$>$~24.50&         &  22.76  & 4.2 &  0.1 &     \\
      &    &          &          &           &$>$~23.50&$>$~24.50&$>$~24.50&$>$~24.00&     &      &     \\ 
  $--$& 45 &    0.15  &    0.33  &  $-$1.42  &         &$>$~22.5*&         &$>$~21.8*&     &      &     \\
   59 &$--$&    0.38  &    0.30  &     0.43  &         &$>$~22.5*&         &$>$~21.8*&     &      &     \\
   60 & 46 &    0.94  &    0.42  &     1.46  & 23.25   &  23.23  &  22.75  &  21.15  & 2.0 &  5.1 & GR  \\
  $--$& 47 & $<~$0.28 &    0.31  &  $<-$0.21 & 22.0*   &  21.9*  &  20.5*  &  19.8*  & 3.6 &  2.8 &     \\
   61 &$--$&    0.31  & $<~$0.20 &  $>$~0.74 & 20.1*   &  20.8*  &  20.1*  &  19.8*  & 0.8 & 31.7 &     \\
   62 & 48 &    0.45  &     0.41 &     0.19  &         &$>$~22.5*&         &$>$~21.8*&     &      &     \\
   63 &$--$&    0.35  & $<~$0.28 &  $>$~0.38 & 22.0*   &  22.2*  &  20.6*  &  20.2*  & 0.3 & 31.7 &     \\
      &    &          &          &           & 21.6*   &  20.1*  &  20.5*  &  19.5*  & 1.5 & 12.8 &     \\
 & & & & \\ \hline \hline
\end{tabular}
\end{minipage}
\end{table*}

A summary of the results of the optical identification and photometry 
for all the 68 radio sources is given in Table~1. The table is arranged
as follows. The source number, their radio flux in the 1.4 and 2.4 GHz catalogs
of Gruppioni et~al. (1997) and the radio spectral index are listed in the
first five columns. The following four columns give respectively the 
$U$, $B$, $V$ and $R$ magnitudes for all the optical counterparts within 5 arcsec
from the radio position. The magnitudes reported in the table 
followed by an asterisk ($\ast$) are those obtained by the photographic plates. 
There are few cases where the $U$ and $B$ magnitudes are
missing; since the size of CCDs in these bands are smaller than in the $V$ and
$R$ ones, the $U$ and $B$ CCD data do not cover the entire area covered by $V$
and $R$ CCD.
Thus for the faintest sources, not visible on the plates, it has not
been possible to measure these magnitudes.
The offset (in arcsec) of the optical counterpart, its likelihood ratio and a note 
about the environment (D=double, GR=group or CL=cluster)
are listed in the last three columns. 

The likelihood ratio technique adopted for source 
identification is the one described by Sutherland \& Saunders (1992), where the likelihood
ratio ($L$) is simply the probability of finding the true optical counterpart in exactly
the position with exactly this magnitude, relative to that of finding a similar chance
background object. As probability distribution of positional errors we adopted a gaussian
distribution with standard deviation of 1.5 arcseconds. This value for $\sigma$ is slightly 
larger than the mean radio positional errors (see Gruppioni et~al. 1997), so to take into 
account the combined effect of radio and optical positional uncertainties. 

For each optical candidate we evaluated also the reliability ($REL$), by
taking into account the presence or absence of other optical candidates for the same radio 
source (Sutherland \& Saunders 1992). Once that $L$ has been computed for all the optical
candidates, one has to choose the best threshold value for $L$ ($L_{th}$) to discriminate 
between 
spurious and real identifications. This is done by studying how the completeness ($C$) and 
reliability ($R$) of the identification sample vary as a function of $L_{th}$. 
The best choice 
for $L_{th}$ is the value which maximizes the estimator $(C+R)/2$. For this purpose, we 
defined $C$ and $R$ as functions of $L_{th}$ according to the formulae given by de Ruiter,
Willis \& Arp (1977). Since we performed our optical identifications on two different
kinds of optical images (with different limiting magnitudes), we had to apply the likelihood
ratio method in two separate steps for two separate sub--samples of our total identification
sample. First, we computed $L$ and $REL$ for each optical counterpart visible on plates ($R 
\simeq 21.8$) within 15 arcseconds from the radio source (we chose a relatively large search 
radius so to obtain a significant statistics for the evaluation of $L$). For this ``bright''
sub--sample thus we computed $C$ and $R$ for different values of $L_{th}$, obtaining as best
values: $L_{th} = 1.5$, $C = 95.7$\%, $R = 89.8$\%. Then we applied the same method
to the optical candidates 
visible only on our CCD exposures, having 21.8 $\lsimeq ~R~ \lsimeq$ 24. For optical 
candidates fainter that $R \simeq 24$ we were not able to give any reliable estimate of $L$ 
and $REL$, since our optical catalogue is fairly incomplete at this limit. 
There are four such faint optical candidates in our identification sample, all 
of them within 
1.5 arcsec from the radio position. Also for the  
``fainter'' identification sub--sample we found $L_{th} = 1.5$ as best choice, with 
corresponding $C = 89.6$\% and $R = 82.0$\%. With this threshold we have
28 likely identifications 
brighter than $R \simeq 21.8$ and 8 with $21.8 \lsimeq R \lsimeq 24$ (plus four additional 
possibly good identifications with objects fainter than $R \sim 24$, too faint
for a reliable determination of their likelihood ratio). The reliability ($REL$)
of each of these optical identifications is always relatively high ($> 80$\%), except for the 
cases where more than one optical candidate with $L > 1.5$ is present for the same source.

As shown in the last column of Table 1,
a significant fraction ($\gsimeq$35\%) of the radio sources with reliable identification
occurs in pairs or groups; moreover many of
these objects show a peculiar optical morphology, suggesting
an enhanced radio emission due to interaction or merging phenomena.
This is in agreement with the results obtained by Kron, Koo \& Windhorst (1985)
and Windhorst et~al. (1995) in the optical identification of sub--mJy, or 
even $\mu$Jy, radio sources. 

\section{Spectroscopy}
\subsection{Observations}
Spectroscopic observations of 34 of the 36 optical 
counterparts with likelihood ratio greater than 1.5 have been carried out
at \eso36 \ telescope.
The sub--sample of these 34 sources is constituted by all
the 28 objects with reliable optical counterpart on the 
photographic plates and by 6 of the 8 with optical 
counterpart on CCDs having $R \leq 23.5$.

The spectroscopic observations have been performed in two different 
observing runs, October 29, 30 and 31 1995, and November 12 1996, with
the EFOSC1 spectrograph (Enard and Delabre, 1992). The spatial scale at~the
detector (RCA 512$^2$, ESO CCD \#8) was 0.61 arcsec~pixel$^{-1}$. 
The spectral ranges covered were usually 3600--7000~\AA \ for the blue objects
at $\sim$6.3~\AA/pix resolution (using the grism B300) and 6000--9200~\AA \
for the red objects at $\sim$7.7~\AA/pix resolution (grism R300).  
In a few cases (for some bright or puzzling objects) we obtained spectra with
both instrumental configuration in order to cover a larger spectral domain.
The slit width was between 1.5 and 2.0 arcsec in order to optimize the balance
of the fraction of object's light within the aperture (due to seeing effects)
and the sky--background  contribution. Because of this relatively small size, 
no effort was made to achieve spectrophotometric precision. 
The exposure times varied from a minimum of 10 minutes for the
brighter optical counterparts, to a maximum of 2 hours for the fainter ones
(with $R$ close to 23). 

\subsection{Data Reduction}
Data reduction has been entirely performed with the NOAO 
``Long--slit'' package in IRAF. 
For every observing night, a bias frame was constructed averaging ten 
``zero exposures'' taken at the beginning and at the end of each night.
The pixel-to-pixel gains were calibrated using flat fields obtained from an
internal quartz lamp. The background was removed subtracting a sky--spectrum
obtained by fitting the intensities measured along the spatial direction in
the column adjacent to the target position. Finally, one--dimensional
spectra were obtained using an optimal extraction (Horne 1986) in order
to have the maximum signal--to--noise ratio also for the fainter objects. 
Wavelength calibration was carried out using Helium--Argon lamps taken at the
beginning and at the end of each night. From the position of the sky lines
in the scientific frames, we estimated the accuracy of the wavelength calibration 
to be about 2~\AA. 
During each night, two standard stars have been observed for flux-calibration purpose.

\subsection{Optical Spectra and Classification}
From our spectroscopic observations we were able to obtain reliable redshifts
for 29 of the 34 observed optical candidates. This corresponds to
$\sim$43\% of the original radio sample. All the 34 spectra are presented in
Figure~2, together with the corresponding optical images, with
superimposed the contour levels of radio emission.

The redshifts have been determined by gaussian--fitting of the emission lines
and via cross--correlation for the absorption--line cases. As templates
for the cross--correlation we used the templates of Kinney et~al. (1996).
These spectra represent a variety of galaxy spectral types --- from
early to late--type and starbursts --- and cover a wide spectral range, from
UV to near--IR, very useful for our sample with a wide range of galaxy
redshifts, up to z~$\approx$~1.

\begin{figure*}
\begin{minipage}{160mm}
\vspace{20cm}
\caption{\label{fig2} EFOSC1 spectra of the 34 spectroscopically observed objects. 
The abscissae are wavelengths in \AA, while the ordinate are monochromatic fluxes in 
arbitrary units.
Below each spectrum, the corresponding $R$ CCD image (when
available, otherwise the $F_K$ photographic plate image) is shown. 
Contour levels of the radio emission  
corresponding to 2,4,6,8,12,15,20,30,50,75,100 $\sigma$ 
are plotted superimposed to each optical image. The size of each image is
1$\times$1 arcmin (in a few cases, where the object was close to the
limit of the CCD, only one of the two dimensions is 1 arcmin) except for
\# 15--10 and \# 38--30, whose images are 1.5$\times$1.5 arcmin because
of the radio emission extent.}
\end{minipage}
\end{figure*}
%
%
%
%
%

The results of spectroscopic analysis are presented in Table~2, which has
the following format: in the first three columns the radio source number (in both
1.4 and 2.4 GHz samples) and the $R$
magnitude are repeated with the same convention as in Table~1. 
The measured redshift, whenever determined, and the list of the 
detected emission lines are in the following columns. In the next two columns
there are the \oii 3727\AA \ ~equivalent width at rest, with associated error, 
followed by the 4000~\AA \ break index\footnote{The 4000~\AA \ ~break index as
defined by Bruzual (1983), is the ratio of the average flux density $f_\nu$
in the bands 4050$\div$4250~\AA \ and 3750$\div$3950~\AA \ at rest.}.
The ``spectral classification'', the ``final classification'' (based also on 
colours) and a short comment are in the last three columns. 
The distinction between spectroscopic types given in column 9 was
based on spectra, continuum indices and visual morphologies. We
divided the objects in several broad classes: Early--type galaxies ({\it Early}),
Late--type galaxies ({\it Late}), AGNs, stars and unclassified objects. 
The classification {\it Late(S)} indicates galaxies in which the detected emission
lines allowed some analysis of line ratios and these ratios are consistent 
with the lines being due to star formation.
Because of the faint magnitudes and the relatively low signal--to--noise ratios
of our spectra, a more detailed spectral classification was very difficult to 
obtain. Moreover, at this stage,
a number of spectra remain unclassified. The final classification reported in column 10
is based also on colours (see next section).

\begin{table*}
 \centering
 \begin{minipage}{170mm}
  \caption{Spectroscopic Results}
 \begin{tabular}{ccrcllrclll}
 & & & & & \\ \hline \hline
 & & & & & \\
\multicolumn{1}{c}{N$_{1.4}$} & 
\multicolumn{1}{c}{N$_{2.4}$} &
\multicolumn{1}{c}{R} &
\multicolumn{1}{c}{z} &
\multicolumn{2}{l}{Emission Lines} &
\multicolumn{1}{c}{W$_0$\oii} &
\multicolumn{1}{c}{D(4000)} &
\multicolumn{1}{l}{Class} &
\multicolumn{1}{l}{Class} &
\multicolumn{1}{l}{Comment} \\
 & & & & \\
 & & & & & \multicolumn{1}{l}{Measured} &\multicolumn{1}{c}{(\AA)}& &\multicolumn{1}{l}{Spectral} 
 &\multicolumn{1}{l}{Final} & \\
 & & & & \\ \hline
 & & & & \\
 03 & 02 & 17.1* & 0.094 & no &                            &              & 2.03 & Early     & Early     & \\
 05 & 03 & 22.03 &   ?   & no &                            &              &      & Early?    & Early     & low S/N\\
 06 &$--$& 16.9* & 0.000 & no &                            &              &      & Star      & Star      & G6 type \\
 07 & 04 & 20.60 &   ?   & no &                            &              &      & BL~Lac?   & BL~Lac?   & \\
 08 &$--$& 19.43 & 2.166 & yes& \lya, ~\civ, ~\ciii        &              &      & AGN1      & AGN1      & x--ray source\\
 09 & 05 & 17.65 & 0.165 & no &                            &              & 2.06 & Early     & Early     & \\
 11 &$--$& 20.9* & 0.368 & yes& \oii, ~\hbe, ~\oiii, ~\hal & 50.8$\pm$8.1 & 1.35 & Late(S)   & Late(S)   & \\
 13 & 08 & 18.8* & 0.229 & yes& \oii, ~\hbe, ~\oiii        & 51.3$\pm$1.3 & 1.32 & Late(S)   & Late(S)   & \\
 15 & 10 & 19.41 & 1.663 & yes& \civ, ~\ciii               &              &      & AGN1      & AGN1      & x--ray source\\
 17 & 12 & 21.7* & 1.147 & yes& \oii                       & 18.5$\pm$0.8 & 1.24 & Unclass.  & Early     & \\
 18 &$--$& 19.2* & 0.209 & yes& \oii, ~\hbe, ~\oiii        & 17.2$\pm$0.5 & 1.47 & Late(S)   & Late(S)   & \\
 20 & 15 & 22.12&$\sim$0.7&no &                            &              &      & Early?    & Early     & z from cluster \\
 22 &$--$& 20.07 & 0.255 & yes& \oii                       & 25.7$\pm$2.0 & 1.67 & Late      & Late      & \\
 24 &$--$& 20.02 & 0.280 & yes& \oii, ~\hbe, ~\oiii        & 40.5$\pm$7.1 & 1.43 & Late(S)   & Late(S)   & \\
 25 & 17 & 20.85 & 0.688 & yes& \oii, ~\oiii               & 18.7$\pm$0.7 & 1.92 & AGN2      & AGN2      & Seyfert~2 gal.\\
 26 & 19 & 20.05 & 0.551 & no &                            &              &$>$1.9& Early     & Early     & \\
 27 & 20 & 17.10 & 0.217 & no &                            &              & 2.04 & Early     & Early     & \\
 30 & 22 & 21.78 & 0.957 & yes& \mgii, ~\oii               &  6.7$\pm$1.2 & 1.75 & Unclass.  & Early     & x--ray source\\
 31 & 23 & 21.58 & 0.757 & no &                            &              & 1.64 & Early     & Early     & \\
 32 & 24 & 22.50 & 0.814 & yes& \oii                       &  9.4$\pm$1.1 & 1.58 & Unclass.  & Early     & \\
 38 & 30 & 19.32 & 0.387 & no &                            &              & 1.97 & Early     & Early     & x--ray source\\
 43 & 34 & 18.60 & 0.219 & no &                            &              & 1.49 & Early     & Early     & noisy spectr. \\
 47 &$--$& 20.05 & 0.579 & no &                            &              &$>$2.1& Early     & Early     & \\
 48 & 36 & 17.40 & 0.154 & yes& \oii, ~\hbe, ~\oiii        &  7.6$\pm$0.9 & 1.61 & Late      & Late      & Bright Spiral \\
 50 &$--$& 18.01 & 0.255 & yes& \oii, ~\hbe, ~\oiii        & 18.7$\pm$0.8 & 1.32 & Late(S)   & Late(S)   & \\
 51 & 38 & 21.90 &   ?   & no &                            &              &      & Early?    & Early     & \\
 52 & 39 & 22.70 & 1.259 & yes& \oii                       &111.4$\pm$9.7 &      & Unclass.  & Early     & \\
 53 & 40 & 21.45 & 0.809 & yes& \oii                       & 28.4$\pm$2.0 & 1.92 & Unclass.  & Early     & merging \\
 55 &$--$& 18.22 & 0.276 & yes& \oii, ~\hal                &  6.9$\pm$0.4 & 1.74 & Late?     & Late?     & \\
 57 & 43 & 22.09 &   ?   &    &                            &              &      & Unclass.  & Late?     & low S/N\\
 60 & 46 & 21.15 & 0.702 & no &                            &              & 2.02 & Early     & Early     & \\
$--$& 47 & 19.8* & 0.275 & yes& \oii, ~\hal                & 11.0$\pm$1.6 & 1.63 & Late?     & Late?     & \\
 61 &$--$& 19.8* & 2.110 & yes& \lya, ~\civ, ~\ciii        &              &      & AGN1      & AGN1      & \\
 63 &$--$& 20.2* & 0.203 & yes& \oii                       & 15.7$\pm$1.6 & 1.18 & Late      & Late      & \\
& & & & \\ \hline \hline
\end{tabular}
\end{minipage}
\end{table*}

A preliminary distinction between different spectroscopic classes was based on
the spectral lines only. Thus, first we divided the
objects into those which show only absorption lines, those which show emission
lines, and those which show no spectral features at all. The ones that show only
absorption lines are most likely to be early--type galaxies. 
For the emission--line objects we attempted a classification separating objects
in which emission lines are probably produced by star--formation, from those in
which an active galactic nucleus is present. 
Four objects have been classified as AGN: three of them have strong broad
lines and unresolved optical image, so they have been classified as QSOs
or type~1 AGN, while the fourth one, with only narrow lines, is likely to be
a type~2 Seyfert galaxy on the basis of the lines intensity ratios. 
In order to produce an objective classification of the narrow 
emission--line objects, we tentatively used the diagnostic diagrams described
by Baldwin, Phillips and Terlevich (1981) and by Rola, Terlevich and Terlevich
(1997), the latter for the higher redshift sources (up to $z \approx 0.7$).
Unfortunately, in a few cases, the observable spectral range accessible 
for high redshift galaxies makes these methods useless,
and the same applies to poor S/N spectra. 
The five spectra allowing the use of diagnostic diagrams (i.e. showing more than one line)
and falling into the H{\tt II}/Starburst region ({\it Late(S)}) show an \oii \ equivalent width 
at rest greater than 15 \AA, suggesting a strong star--formation.
Other five galaxies clearly showing late--type spectra, for which the diagnostic
diagrams could not be applied, have been classified as {\it Late}. They are all at
relatively low redshift ($z < 0.3$) and, on average, have low \oii equivalent
widths. In a couple of cases also H$\alpha$ is detected.
Five emission--line spectra remained unclassified at this stage, because
they could not be unambiguosly classified into any of the above categories 
({\it Early}, {\it Late}, AGN, etc.) on the basis 
of their spectra only. They are all relatively high--$z$ ($>0.8$) galaxies, whose redshift 
determination was mainly based on a single emission line identified with 
[OII]$\lambda$372.
For their classification we used their colours, as well as their absolute magnitudes and
radio luminosities (see next section).

%
For 5 objects, showing no obvious absorption nor
emission lines in their spectra, it was not possible to determine a
redshift, although for one of them (\# 20--15) we assumed a redshift
from nearby galaxies, which are very likely to form a cluster (see notes on
individual sources reported below). 
Four of these objects have $R > 21.8$. Therefore, although we could determine
a redshift for two objects fainter than $R = 21.8$, we can consider this 
magnitude as the approximate limit of our spectroscopic sample.
Three of the objects for which no redshift was determined (\# 05--03, 
\# 20--15, \# 51--38) are tentatively classified as {\it Early} on the basis 
of their red spectra, without prominent emission features.
Instead, \# 57--43, a relatively blue object with
two possible emission lines in its spectrum (not
identified with any obvious spectral feature), remained unclassified at this stage.
The last object (\# 07--04) shows an extremely 
blue spectrum without any distinguishable line or structure, despite the
relatively good S/N.
The spectrum shape, together with its inverted radio spectrum and optical
colours, make it a possible BL Lacertae object. 
%

To summarize the results of our spectral classification, we subdivided the 34 
spectroscopically observed objects into the following populations:
\begin{description}
\item [\bf{12~}] early--type galaxies showing only absorption lines (or no detectable
lines at all, but with red spectra);
\item [\bf{~5~}] late--type objects, with continua typical of evolved star population,
but showing modest \oii \ (and eventually H$\alpha +$ NII) emission lines; 
\item [\bf{~5~}] star--forming emission line galaxies with more than one line in their spectra
and W(\oii) $>$ 15~\AA;
\item [\bf{~5~}] Active Galactic Nuclei, consisting of 3 broad--line QSOs, 1
Seyfert 2 galaxy and a possible BL Lac object;
\item [\bf{~1~}] star;
\item [\bf{~6~}] spectroscopically unclassified objects, 5 of which have a redshift 
determined mainly on the basis of a single
emission line.
\end{description}

\subsection{Notes on individual sources}
Brief comments on the optical and/or radio properties are given for all the objects
in Table 2 and for a few additional sources.

\begin{description}
\item [\bf{03--02}] Bright early--type galaxy, in a group; the close--by galaxy
at $\sim 12^{\prime \prime}$ north of the radio position has the same redshift.
\item [\bf{05--03}] The most likely identification ($L=5.7$) is with a very faint galaxy, 
classified as {\it Early}, 
with a low S/N spectrum for which a redshift determination was not possible.
It is likely to be a member of a compact group of galaxies, since other four objects 
are within 5$^{\prime \prime}$. The brightest
one ($L=4.5$) has a tentative redshift of 0.165, but, due to its much brighter magnitude
with respect to the other members of the group, it is more likely to be a foreground object.
\item [\bf{06--00}] Bright G star. 
\item [\bf{07--04}] Blue object without any obvious spectral feature in its spectrum.
Its colours ($U-B=-0.2$, $B-V=0.56$, $V-R=0.44$),
together with its inverted radio spectrum ($\alpha_r = -0.42$), may suggest that this 
object is a BL Lac AGN. However, this identification has the lowest likelihood ratio
in our sample ($L=1.8$) because of its relatively large distance from the radio position
(3.2 arcsec).
\item [\bf{08--00}] Broad--line quasar (Zitelli et~al. 1992; MZZ7801), which is also 
X--ray emitter.
\item [\bf{09-05}] Bright early--type galaxy with radio emission probably powered by a mini--AGN
in its nucleus, as suggested by its inverted radio spectrum ($\alpha_r = -0.68$).
\item [\bf{11--00}] Blue, emission--line galaxy with line ratios consistent with the
lines being due to star--formation.
\item [\bf{13--08}] Emission--line galaxy with line ratios consistent with the
lines being produced by a strong star--formation activity.
\item [\bf{15--10}] Broad--line quasar (MZZ5571), also X--ray emitter (Zamorani et al. in preparation).
\item [\bf{17--12}] High--$z$ (1.147) galaxy, whose redshift determination is based on the
presence of a single, relatively strong (EW=18.5~\AA) emission--line, identified with \oii$\lambda$3727.
\item [\bf{18--00}] Emission line galaxy with a starburst--like spectrum; the southern 
companion is at the same redshift and shows a late--type galaxy spectrum.
\item [\bf{20--15}] Faint, red galaxy surrounded by a number of galaxies with similar colours.
It is likely to be at the center of a cluster.
The cross--correlation analysis does not lead to any statistically significant redshift
determination. The shape of the continuum at $\sim$8000~\AA \ suggests the possible 
presence of the CaH$+$K break at $z \sim 1.04$, but also this identification is not
supported by any other emission or absorption line in the spectrum. For this reason,
the redshift of this object has not been determined from its spectrum, but it 
has been tentatively estimated from the redshifts we measured for three other cluster 
members ($z \simeq 0.7$). This object has an inverted radio spectrum ($\alpha_r = -0.24$). 
Due to its red colours, we classified this object as an {\it Early} galaxy.
\item [\bf{22--00}] Noisy spectrum, with only a relatively strong \oii \ (EW=25.7~\AA) but
a quite reddened continuum; a brighter galaxy, 19$^{\prime \prime}$ south of the radio 
position has the same redshift. 
\item [\bf{24--00}] Late--type spectrum galaxy with lines produced by strong star--formation 
activity. 
\item [\bf{25--17}] Very red emission--line object classified as Seyfert 2 on the basis of its line--ratios,
while its colours are more typical of an evolved elliptical/S0 galaxy. It has a companion at the 
same redshift within a few arcseconds, so that interaction could be partially responsible for the
enhanced radio and line emission.
\item [\bf{26--19}] Early--type galaxy in a faint, inverted--spectrum
($\alpha_r = -0.11$) radio source.
\item [\bf{27--20}] Bright early--type galaxy at relatively low redshift ($z = 0.217$).
\item [\bf{30--22}] Triple radio source identified with a high redshift galaxy ($z = 0.957$), which is
also an X--ray source. In the UV portion of the spectrum, some possible broad lines (\mgii$\lambda$2798
\cii$\lambda$2326) indicate the presence of an active nucleus, but the continuum 
shortward of 5500~\AA \ is noisy
and shows a suspect fall. Its colours are consistent with this object being an early--type galaxy.
It is surrounded by other fainter objects, suggesting the presence of a group, but all of them are
too faint for any redshift determination.
\item [\bf{31--23}] The faint early--type galaxy suggested as the most likely identification
is the closest to the radio position. An equally
faint, but bluer galaxy with strong \oii \ at the same redshift is at 4.3 arcsec from the 
radio source. Other faint galaxies within a few arcsec suggest the presence of a group.
\item [\bf{32--24}] A compact group of faint galaxies coincides with the radio position. 
The more likely identification is with a high redshift galaxy ($z = 0.814$)
showing a moderate \oii \ emission--line in its spectrum. The colours are
consistent with this object being an early--type galaxy.
\item [\bf{38--30}] Early--type galaxy coincident with the central component of a triple radio source,
which is also an X--ray emitter.
\item [\bf{43--34}] Early--type spectrum galaxy (but noisy spectrum!) with a disky--like morphology.
The spectrum is almost a straight line over the whole observed range and its shape is very
similar to those of the young, reddened galaxies found by Hammer et~al. (1997).
\item [\bf{47--00}] Early--type galaxy at the center of a small group at $z = 0.579$. The object at 5$^{\prime \prime}$ is a blue compact galaxy at the same
redshift.
\item [\bf{48--36}] Bright spiral galaxy, spectrally classified as {\it Late}. Its line--ratios suggest
current star--formation activity, although the spectrum is dominated by old stellar continuum.
\item [\bf{49--37}] A classical double radio source with no obvious optical identification. The CCD image 
suggests a possible association with a faint cluster.
\item [\bf{50--00}] Late--type spectrum galaxy, with line ratios consistent with the lines
being due to star--formation activity.
\item [\bf{51--38}] Faint pair of sources, possibly forming a merging system. Despite a reasonable 
S/N in their spectra it was not possible to identify any obvious structure, nor
obtaining a reliable redshift determination. Due to their
red colours, this pair has been associated to the {\it Early} class. 
\item [\bf{52--39}] Very faint, high--$z$ (1.259) emission line galaxy, whose redshift determination was
based on a single, strong emission--line (EW = 111~\AA), identified with 
\oii$\lambda$3727. The significant detection of continuum shortward of the line seems to exclude the 
\lya \ hypothesis. The colours of this object, together with its radio and optical luminosities are 
consistent with it being a high--$z$ elliptical galaxy.
\item [\bf{53--40}] Very faint, extremely red ($B-R > 3$) \oii \ emitting
galaxy at relatively high $z$ (0.809).
This is probably a close merging system, since its CCD image shows that it has two faint
nuclei. Moreover, it is likely to be surrounded by a faint cluster. Due to its very red colours
and to the inverted spectrum of its radio emission ($\alpha = -0.28$), it is likely
to be an evolved object with a mini--AGN in its nucleus, whose radio and line emission are enhanced
by the on--going merging.     
\item [\bf{55--00}] Late--type, H$\alpha$ emitting galaxy with peculiar optical morphology.
This is a very puzzling objects, since its distorted optical and radio morphologies
strongly suggest this galaxy being in interaction with another close disky--galaxy (optical and
radio tails connecting the two objects are clearly visible). However, their redshifts show a 
significant cosmological distance between them ($\Delta$v = 6534 km/s), which makes the suggested 
interaction rather
unlikely. Moreover, the optical counterpart of the radio source has a close companion at the same 
redshift (at a distance of 6.1 arcsec).
\item [\bf{56--42}] This galaxy, although relatively bright ($R = 22.01$)
and with a high likelihood ratio, has not been 
spectroscopically observed, since it was not in the area covered by the CCD exposures and was close to
the detection limits of the photographic plates. However, it fell by chance in a 5 mins CCD frame 
taken during the last spectroscopic run, thus we could determine its magnitude and position. 
\item [\bf{57--43}] Blue unclassified object. The spectrum shows two possible emission features 
(a large at $\sim$8820~\AA \ and a narrow at $\sim$7250~\AA) which we were not able to identify 
with any obvious spectral line. It is possible
that one of the two lines be a spurious one, since it is a very low S/N spectrum. Due to its colours
and spectral shape this object has been associated to the {\it Late} class. 
\item [\bf{58--44}] The most likely identification for this object is a bright star, but, due to
its likelihood ratio value $< 1.5$ we assumed it as a mis--identification.
\item [\bf{60--46}] Relatively high--$z$ (0.702) early--type galaxy, possibly in a faint group: 
another object at 6.2$^{\prime \prime}$ has the same redshift and several fainter ones are within 
$\sim8^{\prime \prime}$.
\item [\bf{00--47}] Late--type, H$\alpha$ emitting galaxy with a spectrum very similar to that of 
55--00.
\item [\bf{61--00}] Broad--line quasar (MZZ8668).
\item [\bf{63--00}] Complex optical image constituted by two superimposed sources in the plates, 
whose redshifts ($z = 0.203$ and $z = 0.654$) show that they are not related. Both galaxies 
have emission line 
spectra, the most likely identification being with the galaxy at lower redshift and with moderate 
\oii \ emission (EW = 15.7). The other galaxy is bluer, at higher redshift and with 
strong \oii \ line (EW = 33.8).
\end{description}

\section{Radio and Optical Properties of the Faint Radio Galaxy Population}
Colour--redshift diagrams are presented in figure 3, while 
radio spectral index--radio flux, ~radio~flux--optical~magnitude, magnitude--redshift
and radio luminosity--absolute magnitude plots are presented in Figs 4$a$, $b$, $c$,
and $d$. In both figures the objects are plotted with different symbols according to their spectral
classification, while the spectrally unclassified galaxies are represented by a filled dot with either
a circle or a square around, indicating colour classification from figure 3. The dashed
lines in figure 4$b$ represent different values of the radio--to--optical ratio $R$, defined as 
$R = S \times 10^{\frac{(m-12.5)}{2.5}}$, where $S$ is the radio flux in mJy and $m$ is the 
apparent magnitude.
  
For most objects the colours are consistent with their spectral classification (see 
fig. 3). The different classes of objects are discussed individually below.

\begin{figure}
\centerline{
 \psfig{figure=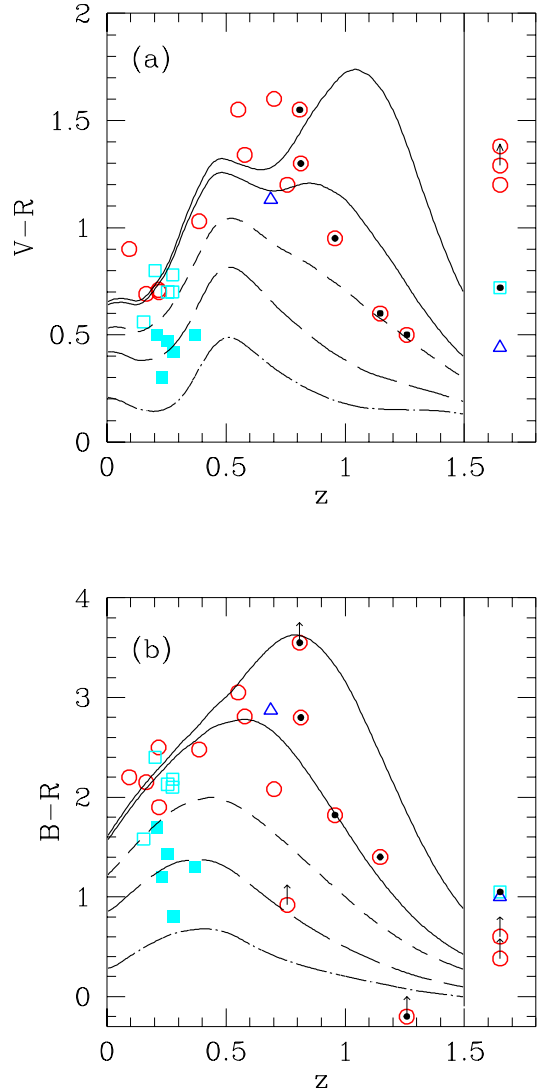,width=17cm}
}
\caption{\label{fig3} $V-R$ ($a$) and $B-R$ ($b$) colour versus redshift
for the extragalactic identifications. The three quasars at $z > 1.5$ are not shown. The 
different symbols represent the different classes of objects: the empty circles stand for 
{\it Early} galaxies, the squares for {\it Late} galaxies (empty for normal and filled for
star--forming galaxies), the empty triangles for AGNs (the type 2 Seyfert galaxy and the
possible BL Lac object)
and the filled dots for spectrally unclassified objects. The latter have 
either a circle or a square around the dot, indicating classification from these diagrams.
At the right side of these figures also the colours for objects without redshift 
determination are plotted.
The different curves correspond to the color--redshift relations 
for galaxies derived from Bruzual \& Charlot (1993) models and represent two different models
for elliptical galaxies (solid lines), Sab--Sbc spirals (dashed line), Scd--Sdm spirals
(long dashed line) and starburst galaxies (dotted--dashed line). The parameters of these 
models are given in Table 3 in Pozzetti et al. (1996).}
\end{figure}

\begin{figure*}
\begin{minipage}{170mm}
\centerline{
 \psfig{figure=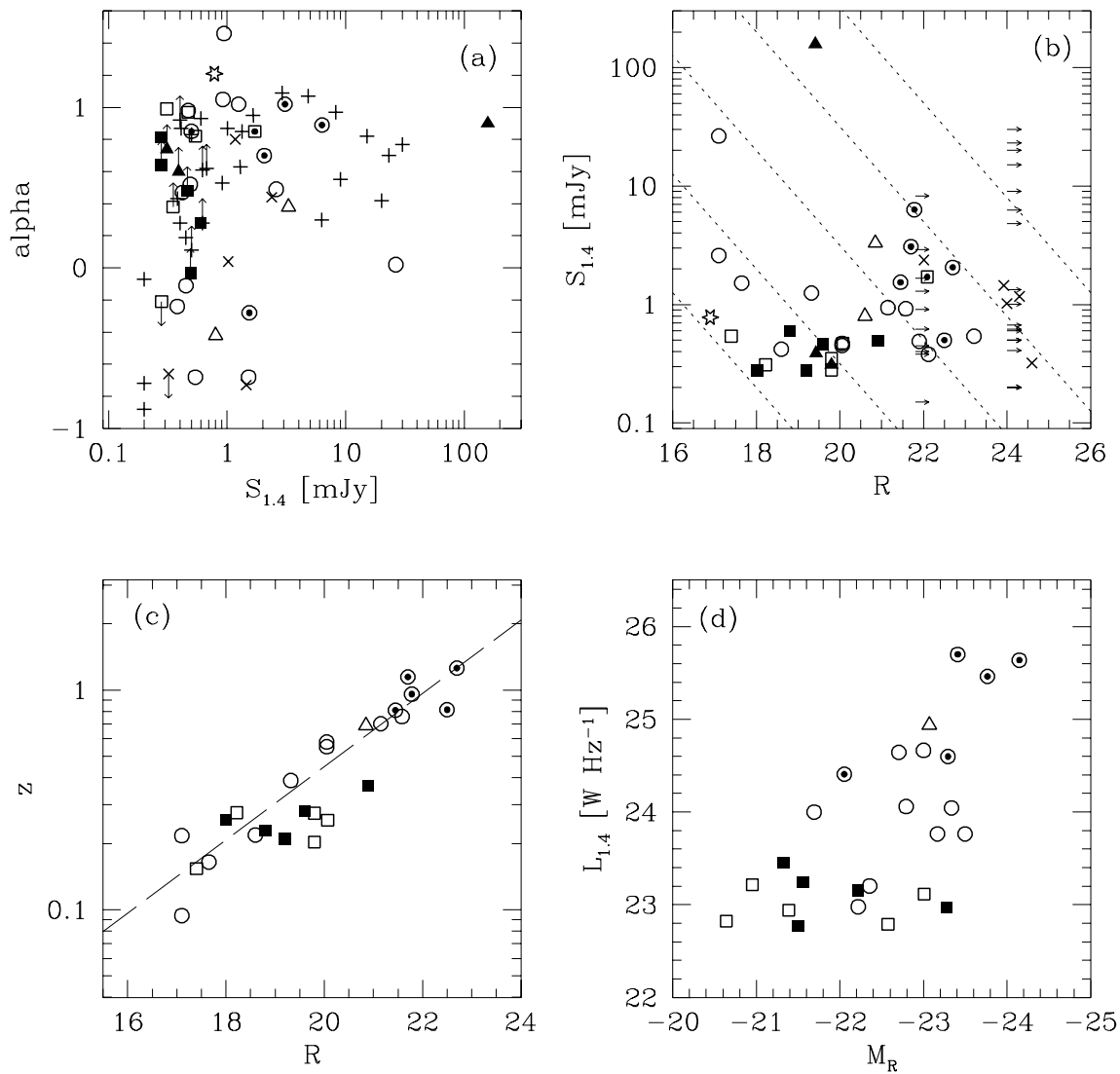,width=17cm}
}
\caption{\label{fig4} Radio spectral index vs. radio flux ($a$) and radio flux vs. $R$ magnitude
($b$) for all radio sources. Symbols are the same as in figure 3, with the addition of filled
triangles for quasars, an empty star for the star, diagonal crosses for objects with optical ID
but no spectrum and vertical crosses for empty fields (arrows in panel $b$).
The dotted lines in panel $b$ are different radio--to--optical ratios, corresponding 
to $logR = 1.5, 2.5, 3.5, 4.5, 5.5$.
Redshift vs. $R$ magnitude ($c$) and radio luminosity vs. absolute $R$ magnitude ($d$) for the
identifications. The dashed line in panel $c$ is the best--fit R--$z$
relation for {\it Early} galaxies in the sample.} 
\end{minipage}
\end{figure*}

\subsection{Early--type galaxies}
In this section we will consider in the group of early--type galaxies both those with
redshift determination and those for which we were unable to measure the redshift
(\# 05--03, \# 20--15 and \# 51--38, plotted in the right side of figure 3), but
with red colours consistent with those of the early--type class. 
The colour--redshift (figs. 3), magnitude--redshift (fig. 4$c$) and radio 
luminosity--absolute magnitude (fig. 4$d$) diagrams for our 
early--type objects (all the empty circles) are consistent with those expected 
for redshifted elliptical and S0 galaxies. 
The radio luminosities for all these galaxies are in the
range $10^{23.0} < P_{1.4~GHz} < 10^{24.8}$ W Hz$^{-1}$ ($H_0 = 50$ km s$^{-1}$ Mpc$^{-1}$,
$q_0 = 0.0$), consistent with them being Fanaroff--Riley I galaxies. 
Consistently with previous results, the early--type galaxies are the dominant population
at $S > 1$ mJy. However, at variance with what found by other authors (see, for example,
Benn et al. 1993), we find a significant number of early--type galaxies also in the
flux range $0.2 \leq S \leq 1$ mJy. A more detailed discussion of the relative
importance of different types of galaxies as a function of radio flux and optical
magnitude will be given in section 6. 

Below 2 mJy, about 13\% of our radiosources have inverted radio spectrum (see figure 4$a$). 
Even if not all of them have been optically identified, it appears that most of these
objects belong to the early--type class, in agreement with the results of
Hammer et~al. (1995), who found an even higher fraction of early--type
galaxies with inverted radio spectra among their identified $\mu$Jy radio sources 
($S_{4.86~GHz} > 16~ \mu$Jy). The suggested
presence of a low--luminosity AGN in the nuclei of these objects, responsible for the observed radio
emission, applies also to the galaxies in our sample, which can be the ``bright'' counterpart
of the Hammer et~al. $\mu$Jy sources (also with very faint optical magnitude, in the
range $23~ \lsimeq ~V~ \lsimeq ~26$). Our non--inverted radio spectra early--type
galaxies are probably powered by an active nucleus, too, since they all have absolute magnitude
greater than $M_R = -21.5$ and relatively high radio luminosity
and other plausible sources of radio emission (HII regions, planetary
nebulae, supernova remnants) cannot account for the observed radio luminosity in galaxies 
brighter than this magnitude (Sadler, Jenkins \& Kotany 1989; Rees 1978; Blanford \& Rees 1978).

\subsection{Late--type and Star--forming Galaxies}
Figures 3 and 4 show that late--type and
star--forming galaxies occupy a narrow range at low--moderate redshift ($0.15 < z < 0.4$) and
most of them have relatively bright apparent magnitudes ($R < 21.0$, $B < 22.5$),
faint radio fluxes ($S_{1.4~GHz} < 1$ mJy) and relatively low radio luminosities 
($L_{1.4~GHz} < 10^{23.5}$ W Hz$^{-1}$).
All the galaxies classified as {\it Late(S)} (i.e. with significant star formation)
on the basis of their spectra have blue colours, consistent with those of late spirals.
The galaxies classified as {\it Late} are, instead, redder and some of them
(\# 55--00, \# 00--47 and \# 63--00) have colours close to the locus expected for
elliptical galaxies. Despite of this, we
will keep for them the spectral classification.

Our {\it Late(S)} galaxies are part of the starburst population at low/intermediate
redshift found in almost all the published sub--mJy identification works (e.g. Windhorst et~al. 1985; 
Benn et~al. 1993), whose radio and optical properties they fully resemble. Also their radio spectral 
indices (which are all steep, see fig.~4$a$) are consistent with their radio emission
being due to syncrothron emission
from supernova remnants, the main source of radio emission at 1.4 GHz in starburst galaxies, with 
typical spectral index in the range $0.5 - 0.9$ ($S_{\nu} \propto \nu^{-\alpha}$). The radio luminosity
of these star--forming galaxies occupy the range $10^{22.7} < L_{1.4~GHz} < 10^{23.5}$ W Hz$^{-1}$,
similar to (but narrower than) that found by Benn et~al. (1993) for their starburst objects. 
The {\it Late} galaxies occupy about the same range of radio and optical luminosities and all but one 
of them have steep radio spectral indices.   

The blue unclassified object \# 57--43 has been associated to the {\it Late} class, due to its colour
properties. However, the radio flux and the apparent magnitude of this object are out of the ranges occupied
by all the other {\it Late} galaxies in our sample. So, its classification remains quite uncertain.

\subsection{AGNs}
Three objects (\# 08--00, \# 15--10 and \# 61--00) have broad emission lines and are in the MZZ quasar 
sample (Zitelli et al. 1992). They are the highest redshift objects in our sample ($z = 2.166, 1.663$ and 2.110 
respectively) and
their absolute optical magnitudes range from $-$25.9 to $-$27.2, all brighter than the limit $M_B = -23$ often
used to separate Seyferts and quasars. They are probably hosted by elliptical galaxies, though only one of them
(15--10, which is also the brightest radio source in our sample) can be defined as radio--loud on the basis of
the ratio between the radio and optical luminosities (Gruppioni and 
Zamorani, in preparation). Their radio powers are in the range 
10$^{25.3} - 10^{27.8}$ W Hz$^{-1}$, comparable to those of lower redshift 3CR and 5C quasars.  

One object (\# 25--17) has only narrow lines and was classified Seyfert 2 on the basis of its emission line ratios.
The radio and optical properties of this object resemble those of the high redshift radio galaxies in our sample 
(described in the following section). In fact, its radio luminosity is 10$^{24.9}$ W Hz$^{-1}$, well inside the
range occupied by our high--$z$ galaxies, and its colours, as well, are as red as those of 
evolved ellipticals. 

\subsection{High--redshift Emission--Line Galaxies}
The group of 5 galaxies at relatively large redshifts ($z > 0.8$) whose spectra were left
unclassified in section 4.3 is composed by intrinsically powerful radio sources.
Their radio powers occupy the range $10^{24.4} < L_{1.4~GHz} < 10^{25.7}$ W Hz$^{-1}$. 
In this range of 
radio powers, far too high for classical late--type and star--forming galaxies, 
Fanaroff--Riley class I
and class II radio galaxies (FRI and FRII) coexist in roughly equal numbers 
(Zirbel and Baum 1995).
All these five radiogalaxies are close to the dividing line between FRI and FRII galaxies in the
radio luminosity--absolute magnitude plane (see, for example, Ledlow and Owen 1996). At the
relatively poor angular resolution of our radio data, a morphological classification is not
possible: three of them are unresolved with a typical upper limit to the size of a few arcsec, 
one consists of a slightly resolved single component, one (\# 30--22) is a triple radio source, 
which, however, does not resemble the classical FRII double radio sources, since the three 
components are not aligned with each other.

Thus, our higher redshift
and more powerful radio sources are very unlikely to be star--forming galaxies as they 
would have been classified on the basis of their emission lines (in most cases just 
one emission line associated with \oii). 
Four out of five of these galaxies have EW([OII]) in the range 6--28~ \AA,
the only exception being \# 52--39, which has an EW larger than 100~ \AA.
Since at the magnitudes and redshifts of these galaxies ~80\% of the
field galaxies have EW([OII])$> 15~$ \AA \ (Hammer et al. 1997),
we conclude that the star formation in these galaxies is not particularly
strong. This would be even more true if some, if not most, of the
emission line flux were due not to stellar but to nuclear ionization.
The relatively low [OII] emission would be consistent with these
radiogalaxies belonging to the FRI class, with \# 52--39 being the only good
candidate for an FRII classification. In fact, for a given absolute
magnitude, line luminosity in FRII radiogalaxies is significantly
higher than in FRI galaxies, while the latter have only slightly higher
line luminosity than normal ``radio quiet'' elliptical galaxies
(Zirbel and Baum 1995).

Figure 3 shows that the colours for these five galaxies are reasonably
consistent with those expected at their redshift for early type galaxies
(ellipticals or early spirals), while they are significantly redder than
those expected for late type spirals or starburst-dominated galaxies.
For this reason, and taking into account also the continuity with
the other early-type galaxies at lower redshift in the redshift--magnitude
plane (figure 4$c$) and their relatively high radio luminosity (figure 4$d$),
we are confident that, despite the presence of [OII] emission in their
spectra, they are physically
unrelated to the star-forming galaxies identified at low-moderate redshift.

However, we can not exclude that some of these red galaxies at relatively
high redshift ($z > 0.8$) are highly obscured galaxies, similar to the
heavily reddened starbursts at $z \leq 1$ recently detected in the 
mid/far--infrared by ISO and in the sub--mm by SCUBA
(Hammer \& Flores 1998; Lilly et al. 1998). These objects, showing star
formation rates, derived from infrared data, in excess of 100 M$_{\odot}$ yr$^{-1}$
are far from being classical star--forming galaxies. Similarly to our red
radio sources, they have red colours and relatively faint [OII] 
emission lines (Hammer et al. 1995). At variance with our high redshift
red galaxies, however, they have smaller radio--to--optical ratios.

\subsection{Radio sources without identification}

For 6 radio sources we have likely optical counterparts, but no
spectroscopic data. Five of them (\# 00--06, \# 21--16, \# 36--27,
\# 41--32 and \# 45--00) are at the limit of our CCD
exposures ($R \sim 24$ and $B \sim 25$), while one (\# 56--42), at 
the limit of our plates, has $R \sim 22$.
Twenty--eight additional radio sources have no optical counterpart either
in the plates (9 objects, $R \geq 21.8$) or in the CCD data (19 objects,
$R \geq 24$). The location of these objects in the radio flux -- optical
magnitude plane is shown in figure 4$b$ (crosses and arrows). Fourteen of 
these objects have $S > 1$ mJy, while 14 have $S < 1$ mJy; most of them have
steep radio spectra ($\alpha > 0.5$). Figure 4$b$ shows that
almost all these objects have a radio to optical ratio significantly higher
than that typical of late type galaxies, including those with
significant star--formation. We therefore conclude that most of them
are likely to be associated with early--type galaxies. This is consistent
with the fact that early--type galaxies constitute the large majority
of the identifications with objects fainter than $B \sim 22.5$. 
In the following discussion we will focus only on the sample of 19  
radio sources without reliable optical 
counterpart on CCD data. Under the assumption that our unidentified
objects are early--type galaxies, we used the $z - R$ magnitude relation 
defined by the objects in figure 4$c$, and other similar relations from
larger samples (Vigotti, private communications, for a sample which 
contains data for about 100 3CR and B3 radiogalaxies) to estimate 
their expected redshifts. The redshift -- magnitude
relation in the $R$ band has a significantly larger scatter than the similar
relation in $K$ band, because the $R$ band, probing the UV rest-frame luminosity,
where a large intrinsic scatter in high--$z$ radiogalaxies is present (Dunlop
et al. 1989), is more affected by possible AGN contamination, recent
star formation or dust. For this reason, given an optical magnitude, only
a relatively large redshift range rather than a true redshift can be estimated. 
For the magnitude corresponding to the limits in 
the CCD data, this range turns out to be $1.2 \lsimeq z \lsimeq 3.0$.
The corresponding radio luminosities would be in the range 
$logP_{1.4~GHz} = 25.6 \pm 0.6$ at $z \sim 2$.

Of course, we can not exclude that a few of the unidentified radio sources
are, instead, associated to starburst galaxies. If so, however, they
should be really extreme objects in terms of their ratio between radio and
optical luminosities. With just one exception (\# 57--43, whose association
with the {\it Late} class is rather uncertain (see Section 4.4)), all the
late--type galaxies in our sample have $1.5~ \lsimeq ~log R~ \lsimeq ~3.0$,
while all but one of the unidentified radio sources have $log R~ \gsimeq ~3.3$.
Radio observations of large and complete samples of spiral galaxies show
that the fraction of such galaxies with $log R$ higher than this
value is $< 10^{-3}$ (Condon 1980; see also Hummel 1981). Given the
number of galaxies with $R \leq 22$ in the entire area covered by our radio
data ($\sim$5400), and assuming that 50\% of them can be considered part of the
starburst class, we can qualitatively estimate that the number of such
galaxies which could have been detected with $log R \geq 3.3$ is at most of the
order of a few (i.e. $< 3$). Obviously, this argument {\it assumes} that the 
radio--to--optical ratio for starburst galaxies does not undergo a strong 
evolution with redshift at $z \gsimeq 1$.

In any case, to shed light on the nature of our unidentified radio sources
deeper optical observations would be needed, but if they indeed are high--$z$
radio galaxies the best observing band would be in the near infrared, where
both the K-correction effects and the dispersion in the redshift--magnitude
relation are much smaller than in the optical. 

\section{Discussion}

In previous works it has been shown that the optically brighter
part of the radio source population at sub--mJy level is
composed largely of starburst galaxies at relatively low redshifts.
Although these results were based on small fraction of spectroscopic
identifications (Benn et al. 1993), it has often been unduly assumed that
they could still be true for the entire population of sub--mJy radio sources.

Our data, based on a significantly higher fraction of optical
identifications (close to 50\%) although on a relatively small sample
of radio sources, do not support this assumption.
In fact, we find that early--type galaxies (including the high--$z$
emission--line galaxies, which are probably the faint end of the
more powerful elliptical radio galaxy population like the 3CR)
are (44 $\pm$ 16)\% of all the radio sources fainter than 1 mJy
identified with galaxies in our sample. In the same radio flux interval
Benn et~al. found a dominance of blue narrow emission line
galaxies and a percentage of early--type galaxies of only about 8\%
(7/84 early--type against 76/84 star--forming galaxies!). The reason of
this discrepancy is very likely to be ascribed to the deeper optical magnitude
reached in our identification work with respect to the previous ones ($B \simeq
24$ to be compared with $B \simeq 22.3$ of Benn et al. 1993). In fact, our
sample suggests an abrupt increase in the fraction of identifications with
early--type galaxies at around $B \simeq 22.5$, which is just above the
magnitude limit of the Benn et al. sample (see figure 4b).
Dividing the sub-mJy sample in two sub--samples (brighter and fainter than
$B = 22.5$), the fraction of early--type galaxies with respect to the total
number of radio sources spectroscopically identified with galaxies increases
from (9 $\pm$ 9)\%, in good agreement with the Benn et al. results
in the same magnitude range, to about 100\%.

Moreover if, as discussed in Section 5.5, also most of the
unidentified radio sources are likely to be associated with high redshift
elliptical radio galaxies, the total fraction of early--type galaxies
in our sub--mJy sample can be estimated to be of the order of
(60 -- 70)\%. This fraction is in good agreement with the prediction
of the model for the evolution of faint radio sources developed by
Rowan--Robinson et al. (1993). Integrating the radio luminosity functions
of spiral galaxies, derived from the Benn et al. sample, and elliptical
galaxies (Dunlop and Peacock 1990) and testing various models for the
evolutionary laws of the spiral luminosity function, they indeed find that
ellipticals still contribute about 60\% of the integrated counts to a
radio limit of 0.1 mJy. Previous models for the interpretation of
the sub--mJy radio counts, based on older luminosity functions and
different models for the evolution, predicted a substantially lower fraction
of early type galaxies in the same flux range (see, for example, Danese et
al. 1987).

Although the percentages of early and late type galaxies we estimated from our
data are in agreement with the predictions of the Rowan--Robinson et al.
models, the redshift distribution of our sample of late type galaxies
(all of them with z $<$ 0.4) appears not to be consistent with the
predictions of the same models, in which a not negligible tail of high
redshift galaxies is expected (see figure 6 in Rowan--Robinson et al. 1993).
Although with relatively large errors, because of the small size of our
sample, the ``local'' volume--densities of our late type galaxies are
consistent with the radio luminosity function of spiral galaxies
computed by Rowan--Robinson et al. (1993). If our conclusion that
most of the unidentified radio sources are likely to be associated with
high redshift elliptical radio galaxies is correct, this would imply
a smaller amount of evolution for the radio luminosity function of late
type galaxies than that assumed in Rowan--Robinson et al. models.
Alternatively, agreement with Rowan--Robinson et al. models could be
obtained only if a significant fraction of our unidentified radio sources
were instead to be classified as starburst galaxies. If actually placed at the
high redshift predicted by these models, their radio powers, in the range 
$10^{24} - 10^{25}$ W Hz$^{-1}$, would require unplausibly high star formation
rates, in excess of a few thousand $M_{\odot} yr^{-1}$, on the basis of the
relation between star formation rate and non--thermal radio emission 
(Condon 1992). Moreover, their radio--to--optical ratios, significantly higher
than those of brighter late type galaxies, would suggest a radio emission
mechanism different from that of local starburst galaxies, probably not
directly related to the star formation episodes. In any case, larger and 
fainter samples of identifications would be needed
in order to choose between these alternatives.

In this respect, it is interesting to compare our results at sub--mJy level
with the existing data at $\mu$Jy level, where the preliminary identification
results obtained for the very few existing samples (Hammer et al. 1995;
Windhorst et al. 1995; Richards et al. 1998) 
are still somewhat unclear. These papers have shown
that the population of $\mu$Jy radio sources is made of star forming
galaxies, ellipticals and AGNs. Given the faint magnitude of the optical
counterparts, the exact fraction of each category is not well defined yet.

\begin{figure}
\centerline{
 \psfig{figure=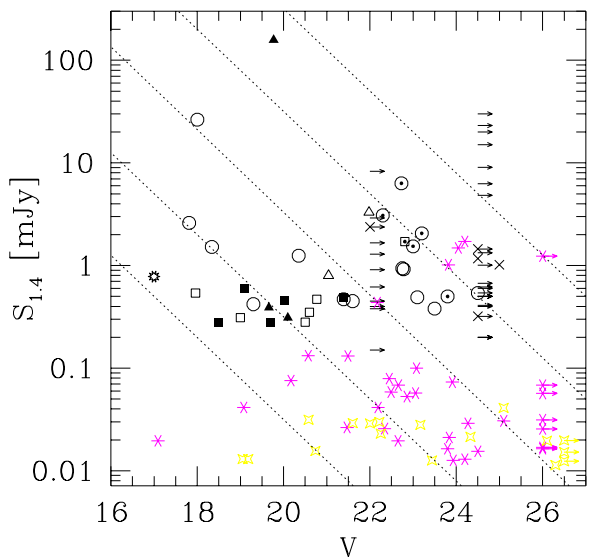,width=9cm}
}
\caption{\label{fig5} 1.4 GHz flux vs. $V$ magnitude for our identifications
(symbols as in fig. 3) and for Hammer et al. (1995) and Windhorst et al.
(1995) counterparts of $\mu$Jy sources (represented, respectively, by 
asterisks and diagonal crosses). The 1.4 GHz fluxes for the
Windhorst et al. objects have been obtained assuming the median spectral index
of the Hammer et al. sample ($\alpha_{med} = 0.2$).
The thick arrows at $V = 26.0$ are the lower limits of Hammer et al., while the
thick arrows at $V = 26.5$ are the lower limits of Windhorst et al.
The dotted lines correspond to the same radio--to--optical ratios as in
figure 4$b$.}
\end{figure}

In figure 5 we show the 1.4 GHz flux versus $V$ magnitude for our data and
the Hammer et al. and Windhorst et al. ones (for which $V$ magnitudes are
available). The 1.4 GHz fluxes have been computed using the 1.5--5 GHz
spectral indices reported by Hammer et al. for their objects and assuming the 
median value of the Hammer et al. spectral index distribution ($\alpha_{med} 
= 0.2$) for the Windhorst et
al. ones. The dotted lines correspond to the same radio--to--optical ratios
as in figure 4$b$. The figure shows that the fraction of radio sources with
large radio--to--optical ratios, typical of the more powerful radio sources,
decreases in the $\mu$Jy samples. For log R $>$ 3.5 this fraction is larger
than 50\% (37/68) in our sample, while it is smaller than 35\% (17/51) for the 
$\mu$Jy samples.
Vice versa, most of the $\mu$Jy radio sources have the same radio--to--optical 
ratio as our low redshift star--forming and elliptical galaxies.

\section{Conclusions}
Optical identifications down to $R \sim 24$ have been performed for a sample 
of 68 radio sources
brighter than 0.2 mJy at 1.4 or 2.4 GHz. About 60\% of the radio sample have a 
likely optical counterpart on deep CCD exposures or ESO plates.
Even in the CCD data, reaching $R \sim 24$ and $B \sim 25$, 19 out of 50 
sources are not identified. 
Spectra have been obtained for 34 optical counterparts brighter than 
$R \simeq 23.5$. The spectra provided enough information 
to determine object type and redshift in most cases (29 objects).
This percentage of spectroscopic identifications is the highest
obtained so far for radio sources in this radio flux range. 
The objects are a mixture of classical early--type galaxies (E/S0), with no
detectable emission lines, covering the redshift range 0.1--0.8, star--forming
and late--type galaxies at moderate redshifts ($z < 0.4$), emission--line
galaxies at relatively high $z$ ($> 0.8$) and AGNs. The star--forming
galaxies are very similar in colour, luminosity and spectral
properties to those found in other sub--mJy surveys (i.e. Benn et al. 
1993). Contrary to previous results, star--forming galaxies do not constitute
the main population in our identification sample. In fact, even at sub--mJy
level the majority of our radio sources are identified with early--type
galaxies. This apparent discrepancy with previous results is due to the
fainter magnitude limit reached in our spectroscopic identifications. In fact, 
the fraction of sub--mJy early--type galaxies in our sample abruptly increases
around $B \simeq 22.5$, which is approximately the magnitude limit reached by
previous identification works. Moreover, the high--$z$ emission--line galaxies 
have spectra, colours, and absolute magnitudes similar to those of
the classical bright elliptical radio galaxies found in surveys carried out at
higher radio fluxes. Their radio luminosity ($10^{24.4} < L_{1.4~GHz} < 10^{25.7}$ 
W Hz$^{-1}$),
far too high for classical star--forming galaxies, is in the range where
FRI and FRII radiogalaxies coexist in roughly equal number. These objects are
therefore likely to constitute the faint radio luminosity end of the distant 
elliptical radio galaxy population, thus further increasing the fraction of
early--type galaxies in our identified sample. Moreover, using mainly the 
large radio--to--optical ratio and the information from the available
limits on the optical magnitudes of the unidentified radio sources, we 
conclude that the great majority of them are likely to be early--type galaxies,
at z $>$ 1. Our classification for these faint objects can be tested
with photometric and spectroscopic observations in the near infrared.
If correct, it would suggest that the evolution of the radio
luminosity function of spiral galaxies, including starbursts, might not be
as strong as suggested in previous evolutionary models.

\section{ACKNOWLEDGEMENTS}

Support for this work was provided by ASI (Italian Space Agency) through 
contracts 95--RS--152 and ARS--96--70.

\label{lastpage}
\end{document}